\documentclass[twocolumn,showpacs,superscriptaddress,prc]{revtex4-2}
\usepackage{dcolumn,pstricks,amssymb}
\usepackage[intlimits]{amsmath}
\ifx\pdfoutput\undefined
\usepackage[dvips]{graphicx}
\else
\usepackage[pdftex]{graphicx}
\usepackage{epstopdf}
\fi
\usepackage{bm}
\usepackage{float}
\usepackage{lipsum}
\usepackage{graphicx}

\begin{document}
\title{\boldmath $\beta$-nuclear-recoil correlation from $^6$He decay in a laser trap }

\author{P.~M\"uller}
\affiliation{Physics Division, Argonne National Laboratory, Lemont, IL 60439, USA}
\author{Y.~Bagdasarova}
 \affiliation{Center for Experimental Nuclear Physics and Astrophysics, University of Washington, Seattle, WA 98105, USA}
\author{R.~Hong}
\affiliation{Center for Experimental Nuclear Physics and Astrophysics, University of Washington, Seattle, WA 98105, USA}
\author{A.~Leredde}
\affiliation{Physics Division, Argonne National Laboratory, Lemont, IL 60439, USA}
\author{K.G.~Bailey}
 \affiliation{Physics Division, Argonne National Laboratory, Lemont, IL 60439, USA}
\author{X.~Fl\'echard}
\affiliation{Normandie Univ, ENSICAEN, UNICAEN, CNRS/IN2P3, LPC Caen, 14000 Caen, France}
\author{A.~Garc\'{\i}a}
\affiliation{Center for Experimental Nuclear Physics and Astrophysics, University of Washington, Seattle, WA 98105, USA}
\author{B.~Graner}
\affiliation{Center for Experimental Nuclear Physics and Astrophysics, University of Washington, Seattle, WA 98105, USA}
\author{A.~Knecht}
\affiliation{Center for Experimental Nuclear Physics and Astrophysics, University of Washington, Seattle, WA 98105, USA}
\affiliation{Paul Scherrer Insitut, 5232 Villigen PSI, Switzerland}
\author{O.~Naviliat-Cuncic}
\affiliation{Normandie Univ, ENSICAEN, UNICAEN, CNRS/IN2P3, LPC Caen, 14000 Caen, France}
\affiliation{National Superconducting Cyclotron Laboratory and Department of Physics and Astronomy, Michigan State University, East Lansing, MI 48824, USA}
\author{T.P.~O'Connor}
\affiliation{Physics Division, Argonne National Laboratory, Lemont, IL 60439, USA}
\author{M.G.~Sternberg}
\affiliation{Center for Experimental Nuclear Physics and Astrophysics, University of Washington, Seattle, WA 98105, USA}
\author{D.W.~Storm}
\affiliation{Center for Experimental Nuclear Physics and Astrophysics, University of Washington, Seattle, WA 98105, USA}
\author{H.E.~Swanson}
\affiliation{Center for Experimental Nuclear Physics and Astrophysics, University of Washington, Seattle, WA 98105, USA}
\author{F.~Wauters}
\affiliation{Center for Experimental Nuclear Physics and Astrophysics, University of Washington, Seattle, WA 98105, USA}
\affiliation{Institut f\"{u}r Kernphysik,  Johannes Gutenberg-Universit\"{a}t Mainz,  55128 Mainz, Germany}
\author{D.W.~Zumwalt}
\affiliation{Center for Experimental Nuclear Physics and Astrophysics, University of Washington, Seattle, WA 98105, USA}

\begin{abstract}
We report the first precise measurement of a $\beta$-recoil correlation from a radioactive noble gas ($^6{\rm He}$) confined via a magneto-optical trap. The measurement is motivated by the search for exotic tensor-type contributions to the charged weak current.
Interpreted as tensor currents with right-handed neutrinos, the measurements yield: $|C_T/C_A|^2\le 0.022$ (90\% C.L.). On the other hand, for left-handed neutrinos the limits are  
$0.007< C_T/C_A <0.111$~(90\%\ C.L.).
The sensitivity of the present measurement is mainly limited by experimental uncertainties in determining the time response properties and the distance between the atom cloud and the micro-channel plate used for recoil ion detection.
\end{abstract}
\maketitle
Precision measurements in nuclear beta decays provide sensitive probes to search for new physics beyond the standard electroweak model (SM) \cite{Gon:2019,Falk:2021}. Few-nucleon systems, like the neutron and light nuclei, offer well controlled environments concerning higher-order contributions such as recoil-order and radiative corrections. These could affect the accuracy in the description of observables and the connection to the relevant physics. High-precision few-body nuclear structure calculations show that these effects can be controlled for light nuclei including the $A=6$ system \cite{Nav:2003, Per:2007, Vai:2009}. Precision measurements in nuclear beta decay enable constraining non-standard model contributions described by the presence of phenomenological scalar and tensor currents. In the absence of deviations from the SM prediction, combined sets of beta-decay observables are particularly helpful to extract both the axial coupling constant in neutron decay, $g_A$, and the up-down quark mixing matrix element $V_{ud}$ free of additional assumptions \cite{Falk:2021}. Furthermore, global analyses of beta-decay data  provide benchmarks for future attempts to achieve significant impacts in the search for new physics, not only in comparison with other precision measurements in beta decay but also relative to current and projected sensitivities at high energies \cite{Gon:2019}.

Pure Gamow-Teller (GT) transitions in light nuclei are one of the cleanest probes for possible tensor contributions \cite{Jack:1957}. 
Early experiments with $^6{\rm He}$~\cite{Joh:1963,Vis:1963} were instrumental to establish the $V-A$ character of the weak interaction \cite{PhysRev.109.193}.
More recently, the first precise measurements of pure GT transitions using ion traps were performed \cite{Fle:2011,Stern:2015,PhysRevC.101.055501}. Data from neutron beta decay, although sensitive to additional degrees of freedom, such as the GT to Fermi ratio, and possible scalar contributions, have improved significantly in recent years \cite{PhysRevC.101.055506,ACORN:2021,UCNA:2020,Sau:2020}.

Here we report a precision measurement of the momentum distribution and beta-ion angular correlation of $^6{\rm Li}$ recoiling ions resulting from the decay of $^6{\rm He}$ atoms confined in a neutral atom laser trap.
The decay-rate function is \cite{Jack:1957,Gon:2016} 
\begin{eqnarray}
\frac{d^2N}{d\Omega_{\theta} dW}
\propto p W q^2
\left(
1 + b \frac{m}{W} + a \frac{p}{W} \cos{\theta}
\right),
\label{eq:ratefunction}
\end{eqnarray}
with $p$, $W$, $m$ the momentum, energy, and mass of the $\beta$, respectively, $q$ the neutrino momentum, and $\theta$ the angle between them. The coefficients $b$ and $a$ can be expressed in terms of the ratio of the tensor to axial couplings \cite{Jack:1957}
\begin{eqnarray}
b \approx {\tilde C_T}+{\tilde C}_T^\prime ~~~~
a \approx -\left(\frac{1}{3}\right) \frac{2-|{\tilde C}_T|^2-|{\tilde C}_T^\prime|^2}{2+|{\tilde C}_T|^2+|{\tilde C}_T^\prime|^2} 
\end{eqnarray}
with ${\tilde C}_T=C_T/C_A.$ The primed and non-primed couplings are such that ${\tilde C}_T=+{\tilde C}_T^\prime$ produces purely left-handed neutrinos while ${\tilde C}_T=-{\tilde C}_T^\prime$ produces purely right-handed neutrinos.

Through the use of a Magneto-Optical Trap (MOT), the $^6{\rm He}$ atoms are well confined and localized in a small ($< 1~\rm{mm}^3$) volume. They are cooled to an absolute temperature of about 1~mK and the recoil ions emerge without any interference from the confining potentials. Laser cooling of alkali metals in MOTs has been exploited for measurements of this kind \cite{Sci:2004,Gor:2005,Vett:2008}. 
Using noble gasses is significantly more challenging because laser-accessible transitions in these elements require excitation to a metastable atomic level, which suffers from a low efficiency. Consequently, the initial production rate of the noble gas radioisotope needs to be proportionally higher.
The noble-gas character, however, also offers advantages in terms of extraction and transport of the radioisotope from the production target. In addition, the 20-eV internal energy of the metastable atom enables determination of crucial MOT parameters via Penning ionization \cite{Mul21}.

\begin{figure}[htbp]
\centering
\includegraphics[width=0.35\textwidth]{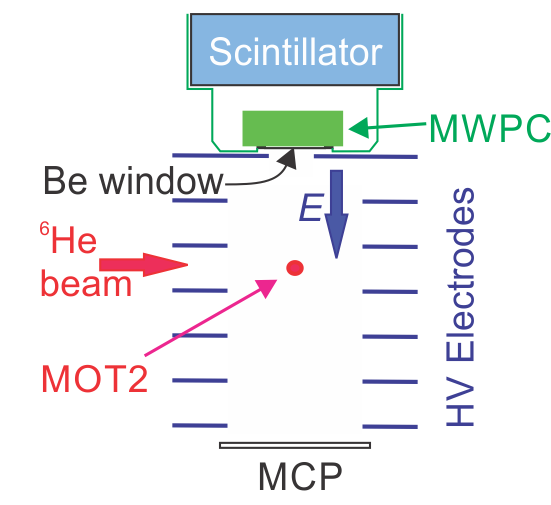}
\caption{Schematic cross section through the decay detection setup surrounding the atomic cloud (MOT2). The MOT2 to MCP distance is $\approx 10$~cm and the sketch is roughly to scale.}
\label{fig:setup}
\end{figure}

Details about the apparatus used in this measurement can be found in Refs.~\cite{Mul21, Hon:2016, Hon:2017, Bag:2019,Hon:2016b}.  
The $^6{\rm He}$ atoms were produced via the $^7{\rm Li}(d,t)^6{\rm He}$ nuclear reaction, bombarding a lithium target with an 18-MeV deuteron beam, delivered by the tandem Van de Graaff accelerator at the University of Washington. With a 14-$\mu{\rm A}$ intense beam on target, about $10^{10}$ $^6$He atoms per second were delivered to the experimental apparatus. An RF-driven plasma discharge in a Xe carrier gas was used to excite a fraction of about $10^{-5}$ of the $^6{\rm He}$ atoms to their metastable electronic state. Subsequently, after transverse-cooling and Zeeman-slower stages, the atoms were trapped in a first MOT (MOT1), from where they were passed in bunches, separated by 0.25~s, to a second MOT (MOT2), with the atomic cloud located at the center of a detection setup (Fig.~\ref{fig:setup}).

The vacuum chambers containing MOT1 and MOT2 were separated by a 25 cm long tube that included a narrow (0.5~cm diameter, 3~cm length) section to minimize diffusion of non-trapped $^6{\rm He}$.
Emphasis was placed on producing a small and stable cloud in MOT2, with the highest stability in the vertical direction to obtain a constant distance to the decay detectors for an accurate determination of the ion kinematics.  
Typically, during data taking, the time average of the number of $^6{\rm He}$ trapped atoms in MOT2 was 1500, distributed in a nearly spherical Gaussian cloud, with a FWHM of $0.5$ mm. The laser-trapping system was designed so that it could be switched from trapping $^6{\rm He}$ to $^4{\rm He}$ within one minute. Every few hours, the laser system was set to trap $^4{\rm He}$ to obtain the atom cloud position via a CCD camera imaging system. The drift in the vertical direction was smaller than 0.2~mm over the data taking period. Careful balance of the trap beam intensity, precise control of the laser frequency shift, and position monitoring of the $^6$He Penning- and photo-ions on the MCP in X- and Y-direction ensured that the trap position difference between the isotopes was well below this limit. Gravitational shift due the difference in isotopic mass is negligible.
The detection system shown in Fig.~\ref{fig:setup} comprises the detector assemblies for beta particles (top) and for recoil ions (bottom). To maximize the detection efficiency, a strong homogeneous electric field was applied between the two detectors, allowing the collection of recoil ions emitted in a solid angle close to $4 \pi$. The detection of the beta particle serves as an accurate reference of the decay time while the time of flight (TOF) and position of the recoil ions on the MCP give access to its momentum components.
The beta detection system consisted of a multi-wire proportional chamber (MWPC) for beta particle direction information, followed by a scintillator, to record the beta energy (Fig.~\ref{fig:setup}). A 0.127-mm thin beryllium window, 38.1 mm in diameter, located 70.9 mm above MOT2,  was used to separate the $2\times 10^{-9}$~Torr vacuum in the MOT2 chamber, from the 1~atm pressure (90\% Ar+10\% CO2) in the MWPC. 
The detection solid angle of the beta telescope was defined by the top electrode, used also for generating the ion-accelerating field, 9.94 mm below the beryllium window. This collimator was made of 2 mm thick stainless steel, enough to stop the highest-energy beta particles. The inner diameter of the collimator was 26 mm, giving a $0.91\%$ detection efficiency. A linear motion feedthrough allowed the insertion of a $^{207}{\rm Bi}$ source at the location of the MOT2 cloud for energy and position calibration of the beta detectors.
We obtain excellent agreement of beta energy spectra recorded with the $^{207}{\rm Bi}$ source with spectra obtained via GEANT4-based simulations, which allows a precise energy calibration over the relevant beta energy range and, in particular, defining the energy threshold \cite{Hon:2016b}.

The ion detector was a micro-channel plate system (MCP) from Photonis,
consisting of 2 MCP’s in the so-called chevron configuration, with 75 mm of active diameter. It was read with two sets of delay-line anode wires from RoentDek, defined to be the $x$ and $y$ axes of the coordinate system. An 88\% open area and $0.050$-mm thick nickel grid mask was located on top of the MCP stack, providing a pattern for calibrating the position response of the MCP \cite{Hon:2016}.

A set of 7 stainless-steel parallel plates, 2-mm thick, was used to define the  electric field, of approximately $1.55 \times 10^5$~V/m for the ``full-field'' configuration and $0.77 \times 10^5$~V/m for the ``low-field'' configuration. In the full-field configuration the ions get a net acceleration up to an energy of approximately 15 keV, significantly larger than their initial kinetic energy at emission, which has a maximum of 1.4 keV. Thus, possible variations of detection efficiency due to initial kinetic energy and angle of incidence are minimized. A HV probe (HV-250 from Computer Power Supply, Inc.) with 0.02\% accuracy and 1:10000 reduction ratio, was used to determine the voltage on the plates. A precision-height gauge was used to determine the plates positions relative to the top plate with an accuracy of $15~\mu{\rm m}$, including uncertainties from thermal expansion.

\begin{figure}[htbp]
\centering
\includegraphics[width=0.45\textwidth]{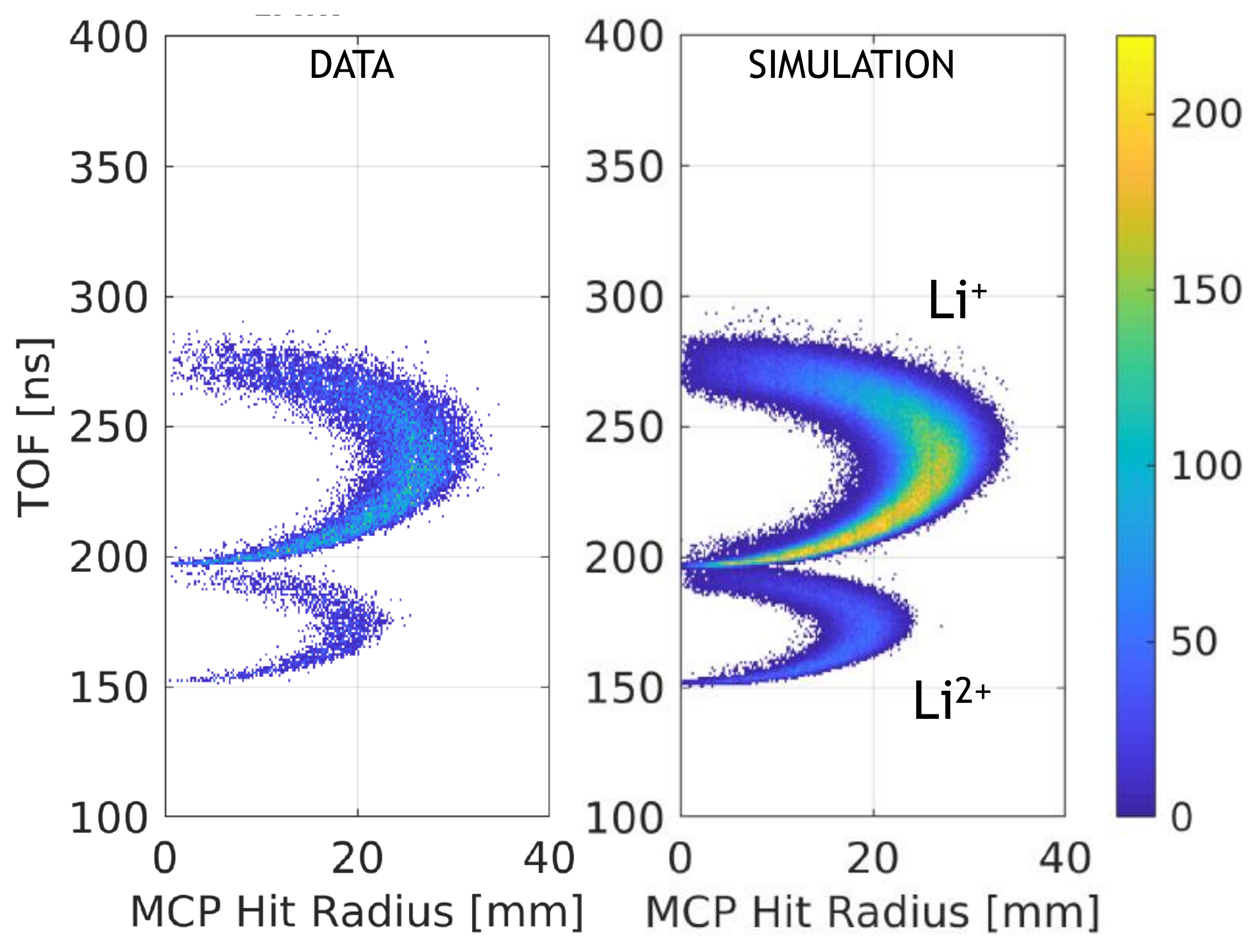}
\caption{Comparing TOF distribution versus hit radius between experiment (left) and GEANT4-based simulations (right) for beta-particle kinetic energy of $1.2 \le K_\beta \le 1.5$~MeV. The two arches in each graph correspond to the two charge states of the Li ions. No ${\rm Li}^{3+}$ ions are observed above background \cite{Hon:2017}.}
\label{fig:TOF-vs-MCP-radius}
\end{figure}

Figure~\ref{fig:TOF-vs-MCP-radius} shows distributions of the TOF versus MCP radius for beta particles with kinetic energies in the range $1.2 \le K_\beta \le 1.5$~MeV. The arches observed for both Li charge states reflect the kinematics. The good agreement between data and simulation gives confidence in the proper description of additional quantities beyond the TOF.

\begin{figure}[htbp]
\centering
\includegraphics[width=0.40\textwidth]{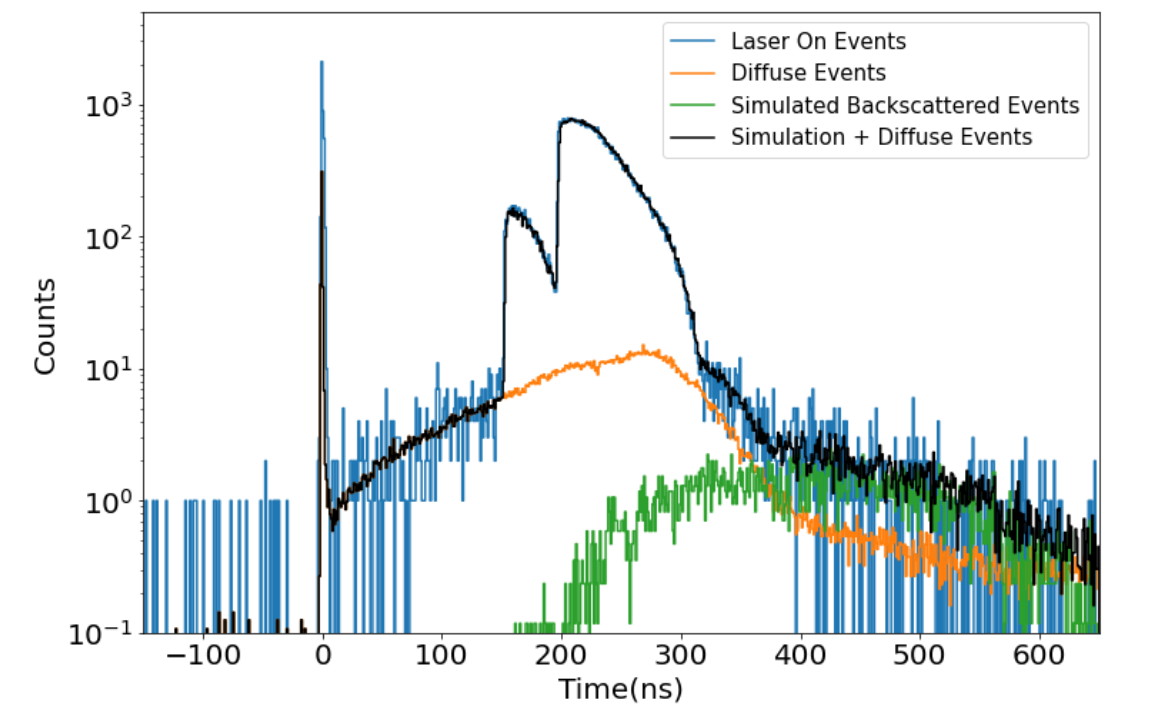}
\hspace*{-0.45cm}\includegraphics[width=0.41\textwidth]{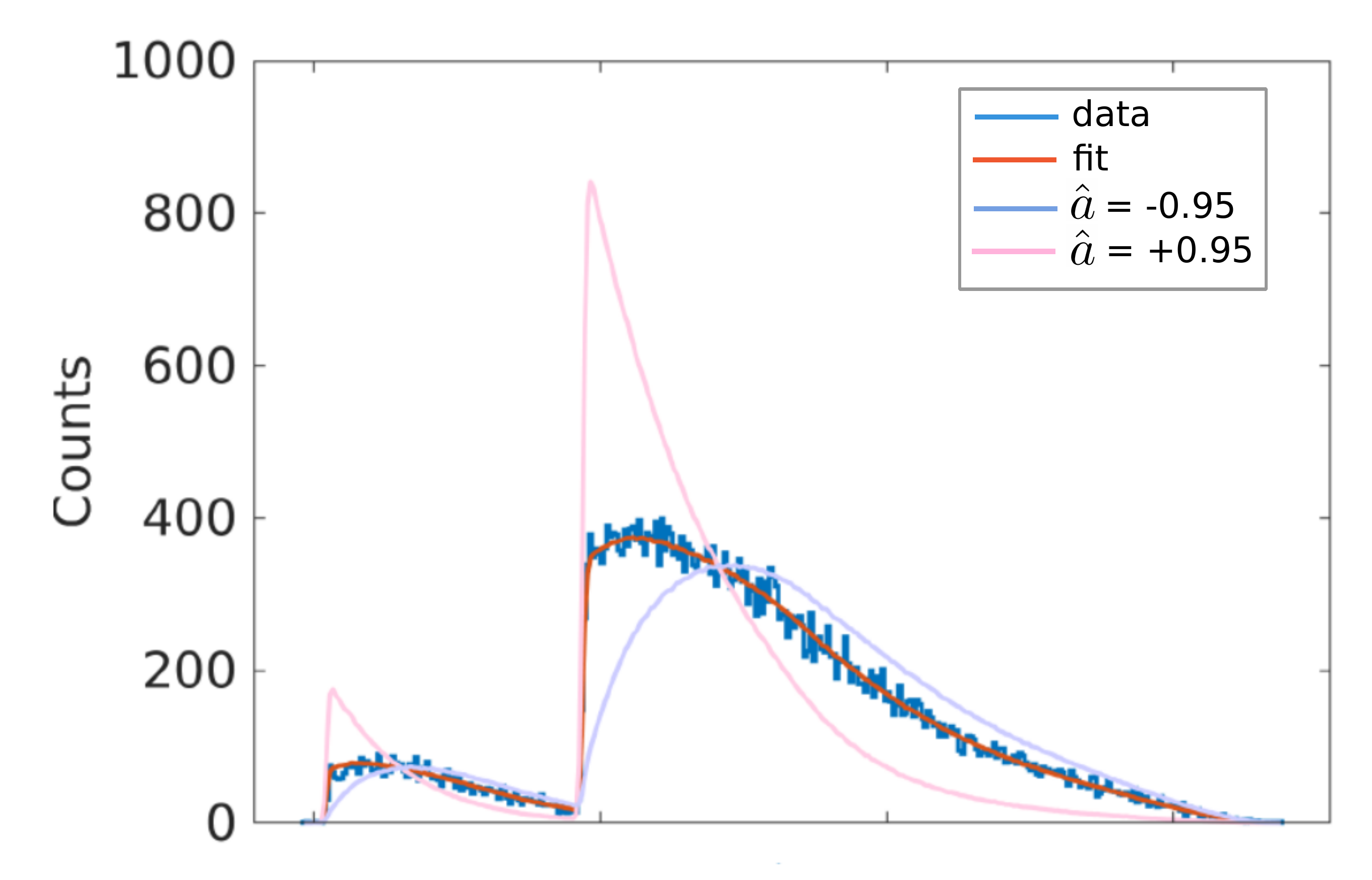}
\hspace*{-0.4cm}\includegraphics[width=0.39\textwidth]{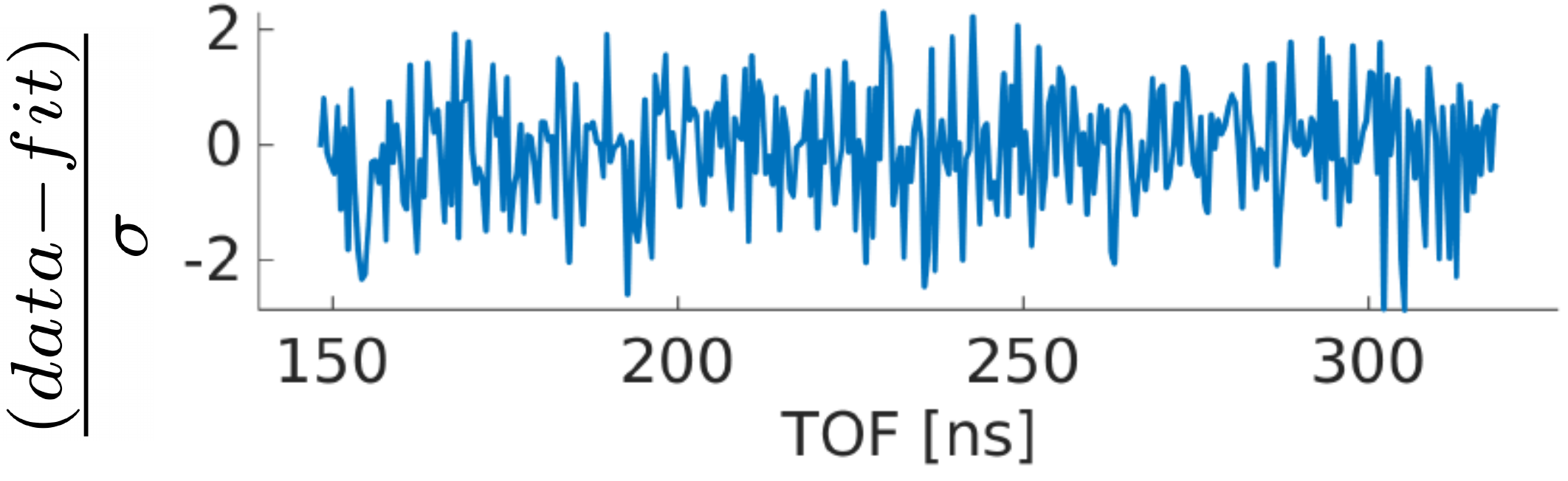}
\caption{Top: TOF spectra under `full field' configuration from laser-on events (blue) and laser-off events (orange) before $Q$-cut. The latter are normalized for times smaller than 147~ns to yield the trapped events spectrum after subtraction. The excess over laser-off events observed at times larger than 350~ns were identified as due to ion backscattering on the MCP. A simulation including the latter, added with the diffuse events, is shown in black. The green curve is the simulation of only the backscattered events. Middle: TOF spectrum and fit after $Q$–cut. Also shown are the $\hat{a} =+0.95$ and $\hat{a}=-0.95$ templates used for the fit. Residuals of fits shown at bottom.}
\label{fig:FFSet1-fits}
\end{figure}

Nearly background-free detection of the $^6{\rm He}$ decay was achieved by requiring a coincidence between the signals from the two MCP anode wires (two for each dimension), the MCP cathode, two MWPC anodes, three MWPC cathodes, and the scintillator. Although attention was placed to minimize diffusion of atoms into the MOT2 chamber, a non-negligible contribution from diffuse $^6{\rm He}$ atoms was observed, which can produce the same coincidence pattern. To determine the contributions from these diffuse $^6{\rm He}$ atoms, data was taken under the same detector settings but where a large amount of $^6{\rm He}$ atoms was purposely injected from the target into the MOT2 chamber via a bypass valve. Figure~\ref{fig:FFSet1-fits} (top panel) shows TOF spectra for diffuse versus trapped events. The diffuse spectrum was normalized for times smaller than 147~ns and subtracted from the trapped spectrum to yield the spectrum from trapped $^6{\rm He}$ only. However, there is a remaining excess of events beyond the maximum TOF of the $^6{\rm Li}^{1+}$ that is not explained by the diffuse data and that had to be further considered. These events, clearly seen in the time interval 450-600~ns in Fig.~\ref{fig:FFSet1-fits} (top panel), are caused by $^6{\rm Li}$ ions that originate in the MOT, scatter back from the Ni grid or the MCP surface, and are only detected on second impact, following their parabolic trajectory in the electric field. The SRIM-2013 software package \cite{Zie:2010,SRIM2013} was used to calculate probabilities, energy, and angular distributions and to obtain the TOF distribution of these backscattered ions (green trace in Fig.~\ref{fig:FFSet1-fits}, top panel). A single scaling fit parameter was used to account for the possibility of the backscattered ion to be neutralized and hence lost to detection. All features of the spectrum are finally very well described by including this contribution in the simulation. 

To reduce the contribution from diffuse $^6{\rm He}$ atoms and suppress ion backscattering events, a condition, called ``$Q$-value cut'', was applied on the sum of the kinetic energies reconstructed from the data under the assumption that the decay occurred in the small MOT volume. Events that originate outside this small volume or include a backscattered ion, lead to non-physical $Q$ values and can be rejected. Once the $Q$-cut is applied, backscattering events in the 450-600~ns region disappear and a large part of the non-trapped $^6{\rm He}$ events is suppressed. The effects of the backscattering in the kinematic region allowed by the $Q$-cut are less obvious. They were explored with the GEANT4-based simulations and were found to be negligible compared to the rest of the systematic uncertainties, discussed below.

To produce the templates for the fits, events were generated assuming the distributions of $^6{\rm He}$ atoms deduced from camera images and photo-ion data. Theoretical details of the decay process that determine the predicted event distributions, such as recoil-order \cite{PhysRevC.12.2016} and radiative corrections \cite{Glu:1998}, were taken into account. The responses of the scintillator, the MWPC and MCP detectors \cite{Hon:2016} were included in the simulations. The ions were tracked via ancillary simulations, performed using field maps calculated using COMSOL$^{\rm{TM}}$ \cite{COMSOL},
and based on detailed measurements of the setup.

Using Monte Carlo simulations, TOF distributions were calculated. For values of $b$ and $a$ close to the Standard Model values, it was shown that making the following replacements in Eq.~(\ref{eq:ratefunction})
\begin{eqnarray}
b &\rightarrow& 0 \nonumber \\
a &\rightarrow&
{\hat a} = a + k \; b
\label{eq:ahat}
\end{eqnarray}
is a proper approximation that yielded results indistinguishable from those obtained using the full Eq.~(\ref{eq:ratefunction}) within the statistical uncertainties \cite{Gon:2016, PhysRevC.101.055501}. 
The coefficient $k$ was found to be strongly dependent on the collection electric field magnitude due to the kinematic coverage and strong correlation between the TOF and the beta-recoil emission angle, with $k=0.031$ for the full-field configuration and $k = 0.059$ for the low-field configuration.
The background-subtracted spectra, like the one shown in Fig.~\ref{fig:FFSet1-fits}, were fit to a linear combination of templates generated using ${\hat a}=-0.95$ and ${\hat a}=0.95$ \footnote{The particular values were chosen for convenience in optimizing the Monte Carlo calculations, but do not affect the results.}. The linear dependence of the rate on $a$, Eq.~(\ref{eq:ratefunction}), allows this method for extraction of a measured value for ${\hat a}$. 

The simulations depend on the distance between the cloud and the MCP, which was fixed using camera images of the $^4{\rm He}$ atom cloud. Other positioning methods are described in Ref.~\cite{Mul21}, based on measuring photo-ions time of flights and varying the electric fields, along with the rationale for choosing the camera images. In addition, it was not possible to independently determine the relative time delays of the beta detector and MCP signal paths with sufficient precision. A parameter describing the time offset, $T_0$, was then left free in the fits of each data set. 

Table~\ref{tab:table1} summarizes the results of the fits. The `Full field 2' data set was not included when quoting the final result below because it showed residuals that were inconsistent with random fluctuations. In particular, the TOF interval 182-190~ns around the leading edge of the Li$^+$ peak, to which the resulting value of $a$ is particularly sensitive (see large difference between fitting templates in Fig.~\ref{fig:FFSet1-fits}) showed $\chi^2/\nu \approx 5.5$. This was interpreted as caused by an unresolved instability of the experimental setup during this run. None of the other data sets showed this type of anomaly.

Table~\ref{tab:table2} lists the main systematic uncertainties. The largest uncertainties arise from the cloud vertical position and the timing resolution. 
The MCP integrated charge and timing delays were found to be position dependent, in a way that could not simply be attributed to amplitude-dependent discriminator issues. The accuracy in determining these parameters was limited by the lack of a calibration system that could faithfully reproduce the conditions of the experiment. Calibrations using $^{249}{\rm Cf}$, which provide $\alpha$-$\gamma$ coincidences, were limited by the $\approx 0.45~\mu{\rm s}$ halflife of the $\gamma$-emitting state and by the qualitative difference between the $\alpha$ energies versus the lithium ions from $^6{\rm He}$. Calibrations using photo-ions were limited to a small central section of the MCP: it was difficult to displace the MOT cloud beyond a couple of mm. Although $\beta$'s that scatter back from the MCP into the scintillator yield coincidences, their timing characteristics could not be assumed to be equal to those for the lithium ions. All these were taken into consideration in determining the uncertainties of the time resolution $\sigma _{T}$. 
The uncertainty in the cloud vertical position reflects position drifts between ${^4}{\rm He}$ calibration runs. More details on systematic uncertainties and potential improvements can be found in Ref. \cite{Bag:2019}.
\begin{table}[htbp]
\caption{\label{tab:table1} Results from the fits to the four independent data sets.}
\begin{ruledtabular}
\begin{tabular}{llllll}
\textrm{Data set} & \hspace{7mm}${\hat a}$ & \hspace{7mm}$T_0$ & ~$\chi^2$& DOF & $p$-value\\
\colrule
Full field 1 & $-0.323(10)$ & $-83.184(30)$ & 343& 340 & 0.45\\
Full field 2 & $-0.311(7)$ & $-83.154(30)$ & 376& 340 & 0.09\\
Low field 1& $-0.319(10)$ & $-83.233(47)$ & 618& 593 & 0.23\\
Low field 2& $-0.331(6)$ & $-83.268(27)$ & 632& 610 & 0.26
\end{tabular}
\end{ruledtabular}
\end{table}

\begin{table}[htbp]
\caption{\label{tab:table2} Main$^a$ systematic evaluated for the high-field case. Values for the low-field case are similar.}
\begin{ruledtabular}
\begin{tabular}{lll}
Parameter& $\delta x$ & $\Delta \hat{a}$ (\%)\\
\colrule
\textrm{Electrode Voltage}&	0.02\%        &  0.16 \\
\textrm{Electrode spacing}&	15 $\mu$m              &  0.27 \\
\textrm{Cloud vertical position}&  200 $\mu$m            &  0.66 \\
Li${2+}$ fraction &  0.30\%                  &  0.30 \\
$\beta$ scattering &  10\%                              &  0.23 \\
Scintillator threshold &	10 keV           &  0.23 \\
Timing resolution, $\sigma_T$   &  50 ps    &  0.80 \\
\hline
\textrm{Total}&		                                           &1.24
\end{tabular}
\end{ruledtabular}
\footnotesize{$^a$ For a more complete list, see Ref.~\protect\cite{Bag:2019}.}\\
\end{table}
The weighted average of the data sets (excluding `Full Field 2') yields
\begin{eqnarray}
{\hat a}= -0.3268\;(46)_{\rm stat}\;(41)_{\rm syst}.
\label{eq:little-a}
\end{eqnarray}

Assuming tensor contributions with right-handed neutrinos ($b=0$ or ${\tilde C}_T=-{\tilde C}_T^\prime$) the result above implies $|{\tilde C}_T|^2\le 0.022$ (90\% C.L.)
On the other hand, assuming purely left-handed neutrinos (${\tilde C}_T=+{\tilde C}_T^\prime$) yields:
\begin{eqnarray}
0.007< {\tilde C}_T <0.111 ~{\rm (90\%\ C.L.).}
\end{eqnarray}
The latter are dominated by the low-field data due to the larger value of the constant $k$ in Eq.~(\ref{eq:ahat}). 

In summary, the results are consistent with the Standard Model. While these limits are not more stringent compared to previous work, such as those in Refs.~\cite{Joh:1963,Fle:2011,Stern:2015,Sau:2020}, they demonstrate the first precise determination of $\beta$-recoil correlations from a neutral atom trap of a noble gas. The work identified sources of limitations that affect similar experiments searching for scalar and tensor currents in nuclear beta decays \cite{Gor:2005,PhysRevLett.120.062502,PhysRevC.101.035501}, studies of atomic shake-off processes \cite{PhysRevA.97.023402}, nuclear spectroscopy \cite{WILSON2021165806} and searches for Dark Matter \cite{HUNTER:2021}.

This work is supported by the Department of Energy, Office of Nuclear Physics, under Contracts No. DE-AC02-06CH11357 and No. DE-FG02-97ER41020, and in part by the US National Science Foundation under Grant No. PHY-11-02511. P.M. acknowledges support through the DOE (SC-NP) Early Career Research Program.
We thank Etienne Li\'enard for help in the early part of this experiment, and Eric Smith, Joben Pederson, and Brittney Dodson for technical support and accelerator operations.

\bibliography{betadecay.bib}

\newcommand{\noopsort}[1]{} \newcommand{\printfirst}[2]{#1}
  \newcommand{\singleletter}[1]{#1} \newcommand{\switchargs}[2]{#2#1}
\begin{thebibliography}{36}%
\makeatletter
\providecommand \@ifxundefined [1]{%
 \@ifx{#1\undefined}
}%
\providecommand \@ifnum [1]{%
 \ifnum #1\expandafter \@firstoftwo
 \else \expandafter \@secondoftwo
 \fi
}%
\providecommand \@ifx [1]{%
 \ifx #1\expandafter \@firstoftwo
 \else \expandafter \@secondoftwo
 \fi
}%
\providecommand \natexlab [1]{#1}%
\providecommand \enquote  [1]{``#1''}%
\providecommand \bibnamefont  [1]{#1}%
\providecommand \bibfnamefont [1]{#1}%
\providecommand \citenamefont [1]{#1}%
\providecommand \href@noop [0]{\@secondoftwo}%
\providecommand \href [0]{\begingroup \@sanitize@url \@href}%
\providecommand \@href[1]{\@@startlink{#1}\@@href}%
\providecommand \@@href[1]{\endgroup#1\@@endlink}%
\providecommand \@sanitize@url [0]{\catcode `\\12\catcode `\$12\catcode
  `\&12\catcode `\#12\catcode `\^12\catcode `\_12\catcode `\%12\relax}%
\providecommand \@@startlink[1]{}%
\providecommand \@@endlink[0]{}%
\providecommand \url  [0]{\begingroup\@sanitize@url \@url }%
\providecommand \@url [1]{\endgroup\@href {#1}{\urlprefix }}%
\providecommand \urlprefix  [0]{URL }%
\providecommand \Eprint [0]{\href }%
\providecommand \doibase [0]{https://doi.org/}%
\providecommand \selectlanguage [0]{\@gobble}%
\providecommand \bibinfo  [0]{\@secondoftwo}%
\providecommand \bibfield  [0]{\@secondoftwo}%
\providecommand \translation [1]{[#1]}%
\providecommand \BibitemOpen [0]{}%
\providecommand \bibitemStop [0]{}%
\providecommand \bibitemNoStop [0]{.\EOS\space}%
\providecommand \EOS [0]{\spacefactor3000\relax}%
\providecommand \BibitemShut  [1]{\csname bibitem#1\endcsname}%
\let\auto@bib@innerbib\@empty
\bibitem [{\citenamefont {González-Alonso}\ \emph {et~al.}(2019)\citenamefont
  {González-Alonso}, \citenamefont {Naviliat-Cuncic},\ and\ \citenamefont
  {Severijns}}]{Gon:2019}%
  \BibitemOpen
  \bibfield  {author} {\bibinfo {author} {\bibfnamefont {M.}~\bibnamefont
  {González-Alonso}}, \bibinfo {author} {\bibfnamefont {O.}~\bibnamefont
  {Naviliat-Cuncic}},\ and\ \bibinfo {author} {\bibfnamefont {N.}~\bibnamefont
  {Severijns}},\ }\href
  {https://doi.org/https://doi.org/10.1016/j.ppnp.2018.08.002} {\bibfield
  {journal} {\bibinfo  {journal} {Progress in Particle and Nuclear Physics}\
  }\textbf {\bibinfo {volume} {104}},\ \bibinfo {pages} {165 } (\bibinfo {year}
  {2019})}\BibitemShut {NoStop}%
\bibitem [{\citenamefont {Falkowski}\ \emph {et~al.}(2021)\citenamefont
  {Falkowski}, \citenamefont {Gonz\'alez-Alonso},\ and\ \citenamefont
  {Naviliat-Cuncic}}]{Falk:2021}%
  \BibitemOpen
  \bibfield  {author} {\bibinfo {author} {\bibfnamefont {A.}~\bibnamefont
  {Falkowski}}, \bibinfo {author} {\bibfnamefont {M.}~\bibnamefont
  {Gonz\'alez-Alonso}},\ and\ \bibinfo {author} {\bibfnamefont
  {O.}~\bibnamefont {Naviliat-Cuncic}},\ }\href
  {https://doi.org/10.1007/JHEP04(2021)126} {\bibfield  {journal} {\bibinfo
  {journal} {Journal of High Energy Physics}\ }\textbf {\bibinfo {volume}
  {2021}},\ \bibinfo {pages} {126} (\bibinfo {year} {2021})}\BibitemShut
  {NoStop}%
\bibitem [{\citenamefont {Navr\'atil}\ and\ \citenamefont
  {Ormand}(2003)}]{Nav:2003}%
  \BibitemOpen
  \bibfield  {author} {\bibinfo {author} {\bibfnamefont {P.}~\bibnamefont
  {Navr\'atil}}\ and\ \bibinfo {author} {\bibfnamefont {W.~E.}\ \bibnamefont
  {Ormand}},\ }\href {https://doi.org/10.1103/PhysRevC.68.034305} {\bibfield
  {journal} {\bibinfo  {journal} {Phys. Rev. C}\ }\textbf {\bibinfo {volume}
  {68}},\ \bibinfo {pages} {034305} (\bibinfo {year} {2003})}\BibitemShut
  {NoStop}%
\bibitem [{\citenamefont {Pervin}\ \emph {et~al.}(2007)\citenamefont {Pervin},
  \citenamefont {Pieper},\ and\ \citenamefont {Wiringa}}]{Per:2007}%
  \BibitemOpen
  \bibfield  {author} {\bibinfo {author} {\bibfnamefont {M.}~\bibnamefont
  {Pervin}}, \bibinfo {author} {\bibfnamefont {S.~C.}\ \bibnamefont {Pieper}},\
  and\ \bibinfo {author} {\bibfnamefont {R.~B.}\ \bibnamefont {Wiringa}},\
  }\href {https://doi.org/10.1103/PhysRevC.76.064319} {\bibfield  {journal}
  {\bibinfo  {journal} {Phys. Rev. C}\ }\textbf {\bibinfo {volume} {76}},\
  \bibinfo {pages} {064319} (\bibinfo {year} {2007})}\BibitemShut {NoStop}%
\bibitem [{\citenamefont {Vaintraub}\ \emph {et~al.}(2009)\citenamefont
  {Vaintraub}, \citenamefont {Barnea},\ and\ \citenamefont {Gazit}}]{Vai:2009}%
  \BibitemOpen
  \bibfield  {author} {\bibinfo {author} {\bibfnamefont {S.}~\bibnamefont
  {Vaintraub}}, \bibinfo {author} {\bibfnamefont {N.}~\bibnamefont {Barnea}},\
  and\ \bibinfo {author} {\bibfnamefont {D.}~\bibnamefont {Gazit}},\ }\href
  {https://doi.org/10.1103/PhysRevC.79.065501} {\bibfield  {journal} {\bibinfo
  {journal} {Phys. Rev. C}\ }\textbf {\bibinfo {volume} {79}},\ \bibinfo
  {pages} {065501} (\bibinfo {year} {2009})}\BibitemShut {NoStop}%
\bibitem [{\citenamefont {Jackson}\ \emph {et~al.}(1957)\citenamefont
  {Jackson}, \citenamefont {Treiman},\ and\ \citenamefont {Wyld}}]{Jack:1957}%
  \BibitemOpen
  \bibfield  {author} {\bibinfo {author} {\bibfnamefont {J.}~\bibnamefont
  {Jackson}}, \bibinfo {author} {\bibfnamefont {S.}~\bibnamefont {Treiman}},\
  and\ \bibinfo {author} {\bibfnamefont {J.~H.}\ \bibnamefont {Wyld}},\
  }\href@noop {} {\bibfield  {journal} {\bibinfo  {journal} {Nuclear Physics}\
  }\textbf {\bibinfo {volume} {4}},\ \bibinfo {pages} {206 } (\bibinfo {year}
  {1957})}\BibitemShut {NoStop}%
\bibitem [{\citenamefont {Johnson}\ \emph {et~al.}(1963)\citenamefont
  {Johnson}, \citenamefont {Pleasonton},\ and\ \citenamefont
  {Carlson}}]{Joh:1963}%
  \BibitemOpen
  \bibfield  {author} {\bibinfo {author} {\bibfnamefont {C.~H.}\ \bibnamefont
  {Johnson}}, \bibinfo {author} {\bibfnamefont {F.}~\bibnamefont
  {Pleasonton}},\ and\ \bibinfo {author} {\bibfnamefont {T.~A.}\ \bibnamefont
  {Carlson}},\ }\href {https://doi.org/10.1103/PhysRev.132.1149} {\bibfield
  {journal} {\bibinfo  {journal} {Phys. Rev.}\ }\textbf {\bibinfo {volume}
  {132}},\ \bibinfo {pages} {1149} (\bibinfo {year} {1963})},\ \bibinfo {note}
  {and see \cite{Glu:1998}}\BibitemShut {NoStop}%
\bibitem [{\citenamefont {Vise}\ and\ \citenamefont {Rustad}(1963)}]{Vis:1963}%
  \BibitemOpen
  \bibfield  {author} {\bibinfo {author} {\bibfnamefont {J.~B.}\ \bibnamefont
  {Vise}}\ and\ \bibinfo {author} {\bibfnamefont {B.~M.}\ \bibnamefont
  {Rustad}},\ }\href {https://doi.org/10.1103/PhysRev.132.2573} {\bibfield
  {journal} {\bibinfo  {journal} {Phys. Rev.}\ }\textbf {\bibinfo {volume}
  {132}},\ \bibinfo {pages} {2573} (\bibinfo {year} {1963})}\BibitemShut
  {NoStop}%
\bibitem [{\citenamefont {Feynman}\ and\ \citenamefont
  {Gell-Mann}(1958)}]{PhysRev.109.193}%
  \BibitemOpen
  \bibfield  {author} {\bibinfo {author} {\bibfnamefont {R.~P.}\ \bibnamefont
  {Feynman}}\ and\ \bibinfo {author} {\bibfnamefont {M.}~\bibnamefont
  {Gell-Mann}},\ }\href {https://doi.org/10.1103/PhysRev.109.193} {\bibfield
  {journal} {\bibinfo  {journal} {Phys. Rev.}\ }\textbf {\bibinfo {volume}
  {109}},\ \bibinfo {pages} {193} (\bibinfo {year} {1958})}\BibitemShut
  {NoStop}%
\bibitem [{\citenamefont {Fl{\'{e}}chard}\ \emph {et~al.}(2011)\citenamefont
  {Fl{\'{e}}chard}, \citenamefont {Velten}, \citenamefont {Li{\'{e}}nard},
  \citenamefont {M{\'{e}}ry}, \citenamefont {Rodr{\'{\i}}guez}, \citenamefont
  {Ban}, \citenamefont {Durand}, \citenamefont {Mauger}, \citenamefont
  {Naviliat-Cuncic},\ and\ \citenamefont {Thomas}}]{Fle:2011}%
  \BibitemOpen
  \bibfield  {author} {\bibinfo {author} {\bibfnamefont {X.}~\bibnamefont
  {Fl{\'{e}}chard}}, \bibinfo {author} {\bibfnamefont {P.}~\bibnamefont
  {Velten}}, \bibinfo {author} {\bibfnamefont {E.}~\bibnamefont
  {Li{\'{e}}nard}}, \bibinfo {author} {\bibfnamefont {A.}~\bibnamefont
  {M{\'{e}}ry}}, \bibinfo {author} {\bibfnamefont {D.}~\bibnamefont
  {Rodr{\'{\i}}guez}}, \bibinfo {author} {\bibfnamefont {G.}~\bibnamefont
  {Ban}}, \bibinfo {author} {\bibfnamefont {D.}~\bibnamefont {Durand}},
  \bibinfo {author} {\bibfnamefont {F.}~\bibnamefont {Mauger}}, \bibinfo
  {author} {\bibfnamefont {O.}~\bibnamefont {Naviliat-Cuncic}},\ and\ \bibinfo
  {author} {\bibfnamefont {J.~C.}\ \bibnamefont {Thomas}},\ }\href
  {https://doi.org/10.1088/0954-3899/38/5/055101} {\bibfield  {journal}
  {\bibinfo  {journal} {Journal of Physics G: Nuclear and Particle Physics}\
  }\textbf {\bibinfo {volume} {38}},\ \bibinfo {pages} {055101} (\bibinfo
  {year} {2011})}\BibitemShut {NoStop}%
\bibitem [{\citenamefont {Sternberg}\ \emph {et~al.}(2015)\citenamefont
  {Sternberg}, \citenamefont {Segel}, \citenamefont {Scielzo}, \citenamefont
  {Savard}, \citenamefont {Clark}, \citenamefont {Bertone}, \citenamefont
  {Buchinger}, \citenamefont {Burkey}, \citenamefont {Caldwell}, \citenamefont
  {Chaudhuri}, \citenamefont {Crawford}, \citenamefont {Deibel}, \citenamefont
  {Greene}, \citenamefont {Gulick}, \citenamefont {Lascar}, \citenamefont
  {Levand}, \citenamefont {Li}, \citenamefont {P\'erez~Galv\'an}, \citenamefont
  {Sharma}, \citenamefont {Van~Schelt}, \citenamefont {Yee},\ and\
  \citenamefont {Zabransky}}]{Stern:2015}%
  \BibitemOpen
  \bibfield  {author} {\bibinfo {author} {\bibfnamefont {M.~G.}\ \bibnamefont
  {Sternberg}}, \bibinfo {author} {\bibfnamefont {R.}~\bibnamefont {Segel}},
  \bibinfo {author} {\bibfnamefont {N.~D.}\ \bibnamefont {Scielzo}}, \bibinfo
  {author} {\bibfnamefont {G.}~\bibnamefont {Savard}}, \bibinfo {author}
  {\bibfnamefont {J.~A.}\ \bibnamefont {Clark}}, \bibinfo {author}
  {\bibfnamefont {P.~F.}\ \bibnamefont {Bertone}}, \bibinfo {author}
  {\bibfnamefont {F.}~\bibnamefont {Buchinger}}, \bibinfo {author}
  {\bibfnamefont {M.}~\bibnamefont {Burkey}}, \bibinfo {author} {\bibfnamefont
  {S.}~\bibnamefont {Caldwell}}, \bibinfo {author} {\bibfnamefont
  {A.}~\bibnamefont {Chaudhuri}}, \bibinfo {author} {\bibfnamefont {J.~E.}\
  \bibnamefont {Crawford}}, \bibinfo {author} {\bibfnamefont {C.~M.}\
  \bibnamefont {Deibel}}, \bibinfo {author} {\bibfnamefont {J.}~\bibnamefont
  {Greene}}, \bibinfo {author} {\bibfnamefont {S.}~\bibnamefont {Gulick}},
  \bibinfo {author} {\bibfnamefont {D.}~\bibnamefont {Lascar}}, \bibinfo
  {author} {\bibfnamefont {A.~F.}\ \bibnamefont {Levand}}, \bibinfo {author}
  {\bibfnamefont {G.}~\bibnamefont {Li}}, \bibinfo {author} {\bibfnamefont
  {A.}~\bibnamefont {P\'erez~Galv\'an}}, \bibinfo {author} {\bibfnamefont
  {K.~S.}\ \bibnamefont {Sharma}}, \bibinfo {author} {\bibfnamefont
  {J.}~\bibnamefont {Van~Schelt}}, \bibinfo {author} {\bibfnamefont {R.~M.}\
  \bibnamefont {Yee}},\ and\ \bibinfo {author} {\bibfnamefont {B.~J.}\
  \bibnamefont {Zabransky}},\ }\href
  {https://doi.org/10.1103/PhysRevLett.115.182501} {\bibfield  {journal}
  {\bibinfo  {journal} {Phys. Rev. Lett.}\ }\textbf {\bibinfo {volume} {115}},\
  \bibinfo {pages} {182501} (\bibinfo {year} {2015})}\BibitemShut {NoStop}%
\bibitem [{\citenamefont {Araujo-Escalona}\ \emph {et~al.}(2020)\citenamefont
  {Araujo-Escalona}, \citenamefont {Atanasov}, \citenamefont {Fl\'echard},
  \citenamefont {Alfaurt}, \citenamefont {Ascher}, \citenamefont {Blank},
  \citenamefont {Daudin}, \citenamefont {Gerbaux}, \citenamefont {Giovinazzo},
  \citenamefont {Gr\'evy}, \citenamefont {Kurtukian-Nieto}, \citenamefont
  {Li\'enard}, \citenamefont {Qu\'em\'ener}, \citenamefont {Severijns},
  \citenamefont {Vanlangendonck}, \citenamefont {Versteegen},\ and\
  \citenamefont {Z\'akouck\'y}}]{PhysRevC.101.055501}%
  \BibitemOpen
  \bibfield  {author} {\bibinfo {author} {\bibfnamefont {V.}~\bibnamefont
  {Araujo-Escalona}}, \bibinfo {author} {\bibfnamefont {D.}~\bibnamefont
  {Atanasov}}, \bibinfo {author} {\bibfnamefont {X.}~\bibnamefont
  {Fl\'echard}}, \bibinfo {author} {\bibfnamefont {P.}~\bibnamefont {Alfaurt}},
  \bibinfo {author} {\bibfnamefont {P.}~\bibnamefont {Ascher}}, \bibinfo
  {author} {\bibfnamefont {B.}~\bibnamefont {Blank}}, \bibinfo {author}
  {\bibfnamefont {L.}~\bibnamefont {Daudin}}, \bibinfo {author} {\bibfnamefont
  {M.}~\bibnamefont {Gerbaux}}, \bibinfo {author} {\bibfnamefont
  {J.}~\bibnamefont {Giovinazzo}}, \bibinfo {author} {\bibfnamefont
  {S.}~\bibnamefont {Gr\'evy}}, \bibinfo {author} {\bibfnamefont
  {T.}~\bibnamefont {Kurtukian-Nieto}}, \bibinfo {author} {\bibfnamefont
  {E.}~\bibnamefont {Li\'enard}}, \bibinfo {author} {\bibfnamefont
  {G.}~\bibnamefont {Qu\'em\'ener}}, \bibinfo {author} {\bibfnamefont
  {N.}~\bibnamefont {Severijns}}, \bibinfo {author} {\bibfnamefont
  {S.}~\bibnamefont {Vanlangendonck}}, \bibinfo {author} {\bibfnamefont
  {M.}~\bibnamefont {Versteegen}},\ and\ \bibinfo {author} {\bibfnamefont
  {D.}~\bibnamefont {Z\'akouck\'y}},\ }\href
  {https://doi.org/10.1103/PhysRevC.101.055501} {\bibfield  {journal} {\bibinfo
   {journal} {Phys. Rev. C}\ }\textbf {\bibinfo {volume} {101}},\ \bibinfo
  {pages} {055501} (\bibinfo {year} {2020})}\BibitemShut {NoStop}%
\bibitem [{\citenamefont {Beck}\ \emph {et~al.}(2020)\citenamefont {Beck},
  \citenamefont {Ayala~Guardia}, \citenamefont {Borg}, \citenamefont
  {Kahlenberg}, \citenamefont {Mu\~noz Horta}, \citenamefont {Schmidt},
  \citenamefont {Wunderle}, \citenamefont {Heil}, \citenamefont {Maisonobe},
  \citenamefont {Simson}, \citenamefont {Soldner}, \citenamefont {Virot},
  \citenamefont {Zimmer}, \citenamefont {Klopf}, \citenamefont {Konrad},
  \citenamefont {Bae\ss{}ler}, \citenamefont {Gl\"uck},\ and\ \citenamefont
  {Schmidt}}]{PhysRevC.101.055506}%
  \BibitemOpen
  \bibfield  {author} {\bibinfo {author} {\bibfnamefont {M.}~\bibnamefont
  {Beck}}, \bibinfo {author} {\bibfnamefont {F.}~\bibnamefont {Ayala~Guardia}},
  \bibinfo {author} {\bibfnamefont {M.}~\bibnamefont {Borg}}, \bibinfo {author}
  {\bibfnamefont {J.}~\bibnamefont {Kahlenberg}}, \bibinfo {author}
  {\bibfnamefont {R.}~\bibnamefont {Mu\~noz Horta}}, \bibinfo {author}
  {\bibfnamefont {C.}~\bibnamefont {Schmidt}}, \bibinfo {author} {\bibfnamefont
  {A.}~\bibnamefont {Wunderle}}, \bibinfo {author} {\bibfnamefont
  {W.}~\bibnamefont {Heil}}, \bibinfo {author} {\bibfnamefont {R.}~\bibnamefont
  {Maisonobe}}, \bibinfo {author} {\bibfnamefont {M.}~\bibnamefont {Simson}},
  \bibinfo {author} {\bibfnamefont {T.}~\bibnamefont {Soldner}}, \bibinfo
  {author} {\bibfnamefont {R.}~\bibnamefont {Virot}}, \bibinfo {author}
  {\bibfnamefont {O.}~\bibnamefont {Zimmer}}, \bibinfo {author} {\bibfnamefont
  {M.}~\bibnamefont {Klopf}}, \bibinfo {author} {\bibfnamefont
  {G.}~\bibnamefont {Konrad}}, \bibinfo {author} {\bibfnamefont
  {S.}~\bibnamefont {Bae\ss{}ler}}, \bibinfo {author} {\bibfnamefont
  {F.}~\bibnamefont {Gl\"uck}},\ and\ \bibinfo {author} {\bibfnamefont
  {U.}~\bibnamefont {Schmidt}},\ }\href
  {https://doi.org/10.1103/PhysRevC.101.055506} {\bibfield  {journal} {\bibinfo
   {journal} {Phys. Rev. C}\ }\textbf {\bibinfo {volume} {101}},\ \bibinfo
  {pages} {055506} (\bibinfo {year} {2020})}\BibitemShut {NoStop}%
\bibitem [{\citenamefont {Hassan}\ \emph {et~al.}(2021)\citenamefont {Hassan},
  \citenamefont {Byron}, \citenamefont {Darius}, \citenamefont {DeAngelis},
  \citenamefont {Wietfeldt}, \citenamefont {Collett}, \citenamefont {Jones},
  \citenamefont {Komives}, \citenamefont {Noid}, \citenamefont {Stephenson},
  \citenamefont {Bateman}, \citenamefont {Dewey}, \citenamefont {Gentile},
  \citenamefont {Mendenhall},\ and\ \citenamefont {Nico}}]{ACORN:2021}%
  \BibitemOpen
  \bibfield  {author} {\bibinfo {author} {\bibfnamefont {M.~T.}\ \bibnamefont
  {Hassan}}, \bibinfo {author} {\bibfnamefont {W.~A.}\ \bibnamefont {Byron}},
  \bibinfo {author} {\bibfnamefont {G.}~\bibnamefont {Darius}}, \bibinfo
  {author} {\bibfnamefont {C.}~\bibnamefont {DeAngelis}}, \bibinfo {author}
  {\bibfnamefont {F.~E.}\ \bibnamefont {Wietfeldt}}, \bibinfo {author}
  {\bibfnamefont {B.}~\bibnamefont {Collett}}, \bibinfo {author} {\bibfnamefont
  {G.~L.}\ \bibnamefont {Jones}}, \bibinfo {author} {\bibfnamefont
  {A.}~\bibnamefont {Komives}}, \bibinfo {author} {\bibfnamefont
  {G.}~\bibnamefont {Noid}}, \bibinfo {author} {\bibfnamefont {E.~J.}\
  \bibnamefont {Stephenson}}, \bibinfo {author} {\bibfnamefont
  {F.}~\bibnamefont {Bateman}}, \bibinfo {author} {\bibfnamefont {M.~S.}\
  \bibnamefont {Dewey}}, \bibinfo {author} {\bibfnamefont {T.~R.}\ \bibnamefont
  {Gentile}}, \bibinfo {author} {\bibfnamefont {M.~P.}\ \bibnamefont
  {Mendenhall}},\ and\ \bibinfo {author} {\bibfnamefont {J.~S.}\ \bibnamefont
  {Nico}},\ }\href {https://doi.org/10.1103/PhysRevC.103.045502} {\bibfield
  {journal} {\bibinfo  {journal} {Phys. Rev. C}\ }\textbf {\bibinfo {volume}
  {103}},\ \bibinfo {pages} {045502} (\bibinfo {year} {2021})}\BibitemShut
  {NoStop}%
\bibitem [{\citenamefont {Sun}\ \emph {et~al.}(2020)\citenamefont {Sun},
  \citenamefont {Adamek}, \citenamefont {Allgeier}, \citenamefont
  {Bagdasarova}, \citenamefont {Berguno}, \citenamefont {Blatnik},
  \citenamefont {Bowles}, \citenamefont {Broussard}, \citenamefont {Brown},
  \citenamefont {Carr}, \citenamefont {Clayton}, \citenamefont {Cude-Woods},
  \citenamefont {Currie}, \citenamefont {Dees}, \citenamefont {Ding},
  \citenamefont {Filippone}, \citenamefont {Garc\'{\i}a}, \citenamefont
  {Geltenbort}, \citenamefont {Hasan}, \citenamefont {Hickerson}, \citenamefont
  {Hoagland}, \citenamefont {Hong}, \citenamefont {Holley}, \citenamefont
  {Ito}, \citenamefont {Knecht}, \citenamefont {Liu}, \citenamefont {Liu},
  \citenamefont {Makela}, \citenamefont {Mammei}, \citenamefont {Martin},
  \citenamefont {Melconian}, \citenamefont {Mendenhall}, \citenamefont {Moore},
  \citenamefont {Morris}, \citenamefont {Nepal}, \citenamefont {Nouri},
  \citenamefont {Pattie}, \citenamefont {P\'erez~G\'alvan}, \citenamefont
  {Phillips}, \citenamefont {Picker}, \citenamefont {Pitt}, \citenamefont
  {Plaster}, \citenamefont {Salvat}, \citenamefont {Saunders}, \citenamefont
  {Sharapov}, \citenamefont {Sjue}, \citenamefont {Slutsky}, \citenamefont
  {Sondheim}, \citenamefont {Swank}, \citenamefont {Tatar}, \citenamefont
  {Vogelaar}, \citenamefont {VornDick}, \citenamefont {Wang}, \citenamefont
  {Wei}, \citenamefont {Wexler}, \citenamefont {Womack}, \citenamefont {Wrede},
  \citenamefont {Young},\ and\ \citenamefont {Zeck}}]{UCNA:2020}%
  \BibitemOpen
  \bibfield  {author} {\bibinfo {author} {\bibfnamefont {X.}~\bibnamefont
  {Sun}}, \bibinfo {author} {\bibfnamefont {E.}~\bibnamefont {Adamek}},
  \bibinfo {author} {\bibfnamefont {B.}~\bibnamefont {Allgeier}}, \bibinfo
  {author} {\bibfnamefont {Y.}~\bibnamefont {Bagdasarova}}, \bibinfo {author}
  {\bibfnamefont {D.~B.}\ \bibnamefont {Berguno}}, \bibinfo {author}
  {\bibfnamefont {M.}~\bibnamefont {Blatnik}}, \bibinfo {author} {\bibfnamefont
  {T.~J.}\ \bibnamefont {Bowles}}, \bibinfo {author} {\bibfnamefont {L.~J.}\
  \bibnamefont {Broussard}}, \bibinfo {author} {\bibfnamefont {M.~A.-P.}\
  \bibnamefont {Brown}}, \bibinfo {author} {\bibfnamefont {R.}~\bibnamefont
  {Carr}}, \bibinfo {author} {\bibfnamefont {S.}~\bibnamefont {Clayton}},
  \bibinfo {author} {\bibfnamefont {C.}~\bibnamefont {Cude-Woods}}, \bibinfo
  {author} {\bibfnamefont {S.}~\bibnamefont {Currie}}, \bibinfo {author}
  {\bibfnamefont {E.~B.}\ \bibnamefont {Dees}}, \bibinfo {author}
  {\bibfnamefont {X.}~\bibnamefont {Ding}}, \bibinfo {author} {\bibfnamefont
  {B.~W.}\ \bibnamefont {Filippone}}, \bibinfo {author} {\bibfnamefont
  {A.}~\bibnamefont {Garc\'{\i}a}}, \bibinfo {author} {\bibfnamefont
  {P.}~\bibnamefont {Geltenbort}}, \bibinfo {author} {\bibfnamefont
  {S.}~\bibnamefont {Hasan}}, \bibinfo {author} {\bibfnamefont {K.~P.}\
  \bibnamefont {Hickerson}}, \bibinfo {author} {\bibfnamefont {J.}~\bibnamefont
  {Hoagland}}, \bibinfo {author} {\bibfnamefont {R.}~\bibnamefont {Hong}},
  \bibinfo {author} {\bibfnamefont {A.~T.}\ \bibnamefont {Holley}}, \bibinfo
  {author} {\bibfnamefont {T.~M.}\ \bibnamefont {Ito}}, \bibinfo {author}
  {\bibfnamefont {A.}~\bibnamefont {Knecht}}, \bibinfo {author} {\bibfnamefont
  {C.-Y.}\ \bibnamefont {Liu}}, \bibinfo {author} {\bibfnamefont
  {J.}~\bibnamefont {Liu}}, \bibinfo {author} {\bibfnamefont {M.}~\bibnamefont
  {Makela}}, \bibinfo {author} {\bibfnamefont {R.}~\bibnamefont {Mammei}},
  \bibinfo {author} {\bibfnamefont {J.~W.}\ \bibnamefont {Martin}}, \bibinfo
  {author} {\bibfnamefont {D.}~\bibnamefont {Melconian}}, \bibinfo {author}
  {\bibfnamefont {M.~P.}\ \bibnamefont {Mendenhall}}, \bibinfo {author}
  {\bibfnamefont {S.~D.}\ \bibnamefont {Moore}}, \bibinfo {author}
  {\bibfnamefont {C.~L.}\ \bibnamefont {Morris}}, \bibinfo {author}
  {\bibfnamefont {S.}~\bibnamefont {Nepal}}, \bibinfo {author} {\bibfnamefont
  {N.}~\bibnamefont {Nouri}}, \bibinfo {author} {\bibfnamefont {R.~W.}\
  \bibnamefont {Pattie}}, \bibinfo {author} {\bibfnamefont {A.}~\bibnamefont
  {P\'erez~G\'alvan}}, \bibinfo {author} {\bibfnamefont {D.~G.}\ \bibnamefont
  {Phillips}}, \bibinfo {author} {\bibfnamefont {R.}~\bibnamefont {Picker}},
  \bibinfo {author} {\bibfnamefont {M.~L.}\ \bibnamefont {Pitt}}, \bibinfo
  {author} {\bibfnamefont {B.}~\bibnamefont {Plaster}}, \bibinfo {author}
  {\bibfnamefont {D.~J.}\ \bibnamefont {Salvat}}, \bibinfo {author}
  {\bibfnamefont {A.}~\bibnamefont {Saunders}}, \bibinfo {author}
  {\bibfnamefont {E.~I.}\ \bibnamefont {Sharapov}}, \bibinfo {author}
  {\bibfnamefont {S.}~\bibnamefont {Sjue}}, \bibinfo {author} {\bibfnamefont
  {S.}~\bibnamefont {Slutsky}}, \bibinfo {author} {\bibfnamefont
  {W.}~\bibnamefont {Sondheim}}, \bibinfo {author} {\bibfnamefont
  {C.}~\bibnamefont {Swank}}, \bibinfo {author} {\bibfnamefont
  {E.}~\bibnamefont {Tatar}}, \bibinfo {author} {\bibfnamefont {R.~B.}\
  \bibnamefont {Vogelaar}}, \bibinfo {author} {\bibfnamefont {B.}~\bibnamefont
  {VornDick}}, \bibinfo {author} {\bibfnamefont {Z.}~\bibnamefont {Wang}},
  \bibinfo {author} {\bibfnamefont {W.}~\bibnamefont {Wei}}, \bibinfo {author}
  {\bibfnamefont {J.~W.}\ \bibnamefont {Wexler}}, \bibinfo {author}
  {\bibfnamefont {T.}~\bibnamefont {Womack}}, \bibinfo {author} {\bibfnamefont
  {C.}~\bibnamefont {Wrede}}, \bibinfo {author} {\bibfnamefont {A.~R.}\
  \bibnamefont {Young}},\ and\ \bibinfo {author} {\bibfnamefont {B.~A.}\
  \bibnamefont {Zeck}} (\bibinfo {collaboration} {UCNA Collaboration}),\ }\href
  {https://doi.org/10.1103/PhysRevC.101.035503} {\bibfield  {journal} {\bibinfo
   {journal} {Phys. Rev. C}\ }\textbf {\bibinfo {volume} {101}},\ \bibinfo
  {pages} {035503} (\bibinfo {year} {2020})}\BibitemShut {NoStop}%
\bibitem [{\citenamefont {Saul}\ \emph {et~al.}(2020)\citenamefont {Saul},
  \citenamefont {Roick}, \citenamefont {Abele}, \citenamefont {Mest},
  \citenamefont {Klopf}, \citenamefont {Petukhov}, \citenamefont {Soldner},
  \citenamefont {Wang}, \citenamefont {Werder},\ and\ \citenamefont
  {M\"arkisch}}]{Sau:2020}%
  \BibitemOpen
  \bibfield  {author} {\bibinfo {author} {\bibfnamefont {H.}~\bibnamefont
  {Saul}}, \bibinfo {author} {\bibfnamefont {C.}~\bibnamefont {Roick}},
  \bibinfo {author} {\bibfnamefont {H.}~\bibnamefont {Abele}}, \bibinfo
  {author} {\bibfnamefont {H.}~\bibnamefont {Mest}}, \bibinfo {author}
  {\bibfnamefont {M.}~\bibnamefont {Klopf}}, \bibinfo {author} {\bibfnamefont
  {A.~K.}\ \bibnamefont {Petukhov}}, \bibinfo {author} {\bibfnamefont
  {T.}~\bibnamefont {Soldner}}, \bibinfo {author} {\bibfnamefont
  {X.}~\bibnamefont {Wang}}, \bibinfo {author} {\bibfnamefont {D.}~\bibnamefont
  {Werder}},\ and\ \bibinfo {author} {\bibfnamefont {B.}~\bibnamefont
  {M\"arkisch}},\ }\href {https://doi.org/10.1103/PhysRevLett.125.112501}
  {\bibfield  {journal} {\bibinfo  {journal} {Phys. Rev. Lett.}\ }\textbf
  {\bibinfo {volume} {125}},\ \bibinfo {pages} {112501} (\bibinfo {year}
  {2020})}\BibitemShut {NoStop}%
\bibitem [{\citenamefont {Gonz\'alez-Alonso}\ and\ \citenamefont
  {Naviliat-Cuncic}(2016)}]{Gon:2016}%
  \BibitemOpen
  \bibfield  {author} {\bibinfo {author} {\bibfnamefont {M.}~\bibnamefont
  {Gonz\'alez-Alonso}}\ and\ \bibinfo {author} {\bibfnamefont {O.}~\bibnamefont
  {Naviliat-Cuncic}},\ }\href {https://doi.org/10.1103/PhysRevC.94.035503}
  {\bibfield  {journal} {\bibinfo  {journal} {Phys. Rev. C}\ }\textbf {\bibinfo
  {volume} {94}},\ \bibinfo {pages} {035503} (\bibinfo {year}
  {2016})}\BibitemShut {NoStop}%
\bibitem [{\citenamefont {Scielzo}\ \emph {et~al.}(2004)\citenamefont
  {Scielzo}, \citenamefont {Freedman}, \citenamefont {Fujikawa},\ and\
  \citenamefont {Vetter}}]{Sci:2004}%
  \BibitemOpen
  \bibfield  {author} {\bibinfo {author} {\bibfnamefont {N.~D.}\ \bibnamefont
  {Scielzo}}, \bibinfo {author} {\bibfnamefont {S.~J.}\ \bibnamefont
  {Freedman}}, \bibinfo {author} {\bibfnamefont {B.~K.}\ \bibnamefont
  {Fujikawa}},\ and\ \bibinfo {author} {\bibfnamefont {P.~A.}\ \bibnamefont
  {Vetter}},\ }\href {https://doi.org/10.1103/PhysRevLett.93.102501} {\bibfield
   {journal} {\bibinfo  {journal} {Phys. Rev. Lett.}\ }\textbf {\bibinfo
  {volume} {93}},\ \bibinfo {pages} {102501} (\bibinfo {year}
  {2004})}\BibitemShut {NoStop}%
\bibitem [{\citenamefont {Gorelov}\ \emph {et~al.}(2005)\citenamefont
  {Gorelov}, \citenamefont {Melconian}, \citenamefont {Alford}, \citenamefont
  {Ashery}, \citenamefont {Ball}, \citenamefont {Behr}, \citenamefont
  {Bricault}, \citenamefont {D'Auria}, \citenamefont {Deutsch}, \citenamefont
  {Dilling}, \citenamefont {Dombsky}, \citenamefont {Dub\'e}, \citenamefont
  {Fingler}, \citenamefont {Giesen}, \citenamefont {Gl\"uck}, \citenamefont
  {Gu}, \citenamefont {H\"ausser}, \citenamefont {Jackson}, \citenamefont
  {Jennings}, \citenamefont {Pearson}, \citenamefont {Stocki}, \citenamefont
  {Swanson},\ and\ \citenamefont {Trinczek}}]{Gor:2005}%
  \BibitemOpen
  \bibfield  {author} {\bibinfo {author} {\bibfnamefont {A.}~\bibnamefont
  {Gorelov}}, \bibinfo {author} {\bibfnamefont {D.}~\bibnamefont {Melconian}},
  \bibinfo {author} {\bibfnamefont {W.~P.}\ \bibnamefont {Alford}}, \bibinfo
  {author} {\bibfnamefont {D.}~\bibnamefont {Ashery}}, \bibinfo {author}
  {\bibfnamefont {G.}~\bibnamefont {Ball}}, \bibinfo {author} {\bibfnamefont
  {J.~A.}\ \bibnamefont {Behr}}, \bibinfo {author} {\bibfnamefont {P.~G.}\
  \bibnamefont {Bricault}}, \bibinfo {author} {\bibfnamefont {J.~M.}\
  \bibnamefont {D'Auria}}, \bibinfo {author} {\bibfnamefont {J.}~\bibnamefont
  {Deutsch}}, \bibinfo {author} {\bibfnamefont {J.}~\bibnamefont {Dilling}},
  \bibinfo {author} {\bibfnamefont {M.}~\bibnamefont {Dombsky}}, \bibinfo
  {author} {\bibfnamefont {P.}~\bibnamefont {Dub\'e}}, \bibinfo {author}
  {\bibfnamefont {J.}~\bibnamefont {Fingler}}, \bibinfo {author} {\bibfnamefont
  {U.}~\bibnamefont {Giesen}}, \bibinfo {author} {\bibfnamefont
  {F.}~\bibnamefont {Gl\"uck}}, \bibinfo {author} {\bibfnamefont
  {S.}~\bibnamefont {Gu}}, \bibinfo {author} {\bibfnamefont {O.}~\bibnamefont
  {H\"ausser}}, \bibinfo {author} {\bibfnamefont {K.~P.}\ \bibnamefont
  {Jackson}}, \bibinfo {author} {\bibfnamefont {B.~K.}\ \bibnamefont
  {Jennings}}, \bibinfo {author} {\bibfnamefont {M.~R.}\ \bibnamefont
  {Pearson}}, \bibinfo {author} {\bibfnamefont {T.~J.}\ \bibnamefont {Stocki}},
  \bibinfo {author} {\bibfnamefont {T.~B.}\ \bibnamefont {Swanson}},\ and\
  \bibinfo {author} {\bibfnamefont {M.}~\bibnamefont {Trinczek}},\ }\href
  {https://doi.org/10.1103/PhysRevLett.94.142501} {\bibfield  {journal}
  {\bibinfo  {journal} {Phys. Rev. Lett.}\ }\textbf {\bibinfo {volume} {94}},\
  \bibinfo {pages} {142501} (\bibinfo {year} {2005})}\BibitemShut {NoStop}%
\bibitem [{\citenamefont {Vetter}\ \emph {et~al.}(2008)\citenamefont {Vetter},
  \citenamefont {Abo-Shaeer}, \citenamefont {Freedman},\ and\ \citenamefont
  {Maruyama}}]{Vett:2008}%
  \BibitemOpen
  \bibfield  {author} {\bibinfo {author} {\bibfnamefont {P.~A.}\ \bibnamefont
  {Vetter}}, \bibinfo {author} {\bibfnamefont {J.~R.}\ \bibnamefont
  {Abo-Shaeer}}, \bibinfo {author} {\bibfnamefont {S.~J.}\ \bibnamefont
  {Freedman}},\ and\ \bibinfo {author} {\bibfnamefont {R.}~\bibnamefont
  {Maruyama}},\ }\href {https://doi.org/10.1103/PhysRevC.77.035502} {\bibfield
  {journal} {\bibinfo  {journal} {Phys. Rev. C}\ }\textbf {\bibinfo {volume}
  {77}},\ \bibinfo {pages} {035502} (\bibinfo {year} {2008})}\BibitemShut
  {NoStop}%
\bibitem [{\citenamefont {Leredde}\ \emph {et~al.}(2021)\citenamefont
  {Leredde}, \citenamefont {Bagdasarova}, \citenamefont {Bailey}, \citenamefont
  {Flechard}, \citenamefont {Garcia}, \citenamefont {Graner}, \citenamefont
  {Hong}, \citenamefont {Knecht}, \citenamefont {Lienard}, \citenamefont
  {Mueller}, \citenamefont {Naviliat-Cuncic}, \citenamefont {O’Connor},
  \citenamefont {Sternberg}, \citenamefont {Swanson}, \citenamefont {Wauters},\
  and\ \citenamefont {Zumwalt}}]{Mul21}%
  \BibitemOpen
  \bibfield  {author} {\bibinfo {author} {\bibfnamefont {A.}~\bibnamefont
  {Leredde}}, \bibinfo {author} {\bibfnamefont {Y.}~\bibnamefont
  {Bagdasarova}}, \bibinfo {author} {\bibfnamefont {K.}~\bibnamefont {Bailey}},
  \bibinfo {author} {\bibfnamefont {X.}~\bibnamefont {Flechard}}, \bibinfo
  {author} {\bibfnamefont {A.}~\bibnamefont {Garcia}}, \bibinfo {author}
  {\bibfnamefont {B.}~\bibnamefont {Graner}}, \bibinfo {author} {\bibfnamefont
  {R.}~\bibnamefont {Hong}}, \bibinfo {author} {\bibfnamefont {A.}~\bibnamefont
  {Knecht}}, \bibinfo {author} {\bibfnamefont {E.}~\bibnamefont {Lienard}},
  \bibinfo {author} {\bibfnamefont {P.}~\bibnamefont {Mueller}}, \bibinfo
  {author} {\bibfnamefont {O.}~\bibnamefont {Naviliat-Cuncic}}, \bibinfo
  {author} {\bibfnamefont {T.}~\bibnamefont {O’Connor}}, \bibinfo {author}
  {\bibfnamefont {M.}~\bibnamefont {Sternberg}}, \bibinfo {author}
  {\bibfnamefont {H.}~\bibnamefont {Swanson}}, \bibinfo {author} {\bibfnamefont
  {F.}~\bibnamefont {Wauters}},\ and\ \bibinfo {author} {\bibfnamefont
  {D.}~\bibnamefont {Zumwalt}},\ }\href {https://doi.org/xxx} {\bibfield
  {journal} {\bibinfo  {journal} {To be submitted}\ } (\bibinfo {year}
  {2022})}\BibitemShut {NoStop}%
\bibitem [{\citenamefont {Hong}\ \emph {et~al.}(2016)\citenamefont {Hong},
  \citenamefont {Leredde}, \citenamefont {Bagdasarova}, \citenamefont
  {Fléchard}, \citenamefont {García}, \citenamefont {Müller}, \citenamefont
  {Knecht}, \citenamefont {Liénard}, \citenamefont {Kossin}, \citenamefont
  {Sternberg}, \citenamefont {Swanson},\ and\ \citenamefont
  {Zumwalt}}]{Hon:2016}%
  \BibitemOpen
  \bibfield  {author} {\bibinfo {author} {\bibfnamefont {R.}~\bibnamefont
  {Hong}}, \bibinfo {author} {\bibfnamefont {A.}~\bibnamefont {Leredde}},
  \bibinfo {author} {\bibfnamefont {Y.}~\bibnamefont {Bagdasarova}}, \bibinfo
  {author} {\bibfnamefont {X.}~\bibnamefont {Fléchard}}, \bibinfo {author}
  {\bibfnamefont {A.}~\bibnamefont {García}}, \bibinfo {author} {\bibfnamefont
  {P.}~\bibnamefont {Müller}}, \bibinfo {author} {\bibfnamefont
  {A.}~\bibnamefont {Knecht}}, \bibinfo {author} {\bibfnamefont
  {E.}~\bibnamefont {Liénard}}, \bibinfo {author} {\bibfnamefont
  {M.}~\bibnamefont {Kossin}}, \bibinfo {author} {\bibfnamefont
  {M.}~\bibnamefont {Sternberg}}, \bibinfo {author} {\bibfnamefont
  {H.}~\bibnamefont {Swanson}},\ and\ \bibinfo {author} {\bibfnamefont
  {D.}~\bibnamefont {Zumwalt}},\ }\href
  {https://doi.org/https://doi.org/10.1016/j.nima.2016.08.024} {\bibfield
  {journal} {\bibinfo  {journal} {Nuclear Instruments and Methods in Physics
  Research Section A: Accelerators, Spectrometers, Detectors and Associated
  Equipment}\ }\textbf {\bibinfo {volume} {835}},\ \bibinfo {pages} {42}
  (\bibinfo {year} {2016})}\BibitemShut {NoStop}%
\bibitem [{\citenamefont {Hong}\ \emph {et~al.}(2017)\citenamefont {Hong},
  \citenamefont {Leredde}, \citenamefont {Bagdasarova}, \citenamefont
  {Fl\'echard}, \citenamefont {Garc\'{\i}a}, \citenamefont {Knecht},
  \citenamefont {M\"uller}, \citenamefont {Naviliat-Cuncic}, \citenamefont
  {Pedersen}, \citenamefont {Smith}, \citenamefont {Sternberg}, \citenamefont
  {Storm}, \citenamefont {Swanson}, \citenamefont {Wauters},\ and\
  \citenamefont {Zumwalt}}]{Hon:2017}%
  \BibitemOpen
  \bibfield  {author} {\bibinfo {author} {\bibfnamefont {R.}~\bibnamefont
  {Hong}}, \bibinfo {author} {\bibfnamefont {A.}~\bibnamefont {Leredde}},
  \bibinfo {author} {\bibfnamefont {Y.}~\bibnamefont {Bagdasarova}}, \bibinfo
  {author} {\bibfnamefont {X.}~\bibnamefont {Fl\'echard}}, \bibinfo {author}
  {\bibfnamefont {A.}~\bibnamefont {Garc\'{\i}a}}, \bibinfo {author}
  {\bibfnamefont {A.}~\bibnamefont {Knecht}}, \bibinfo {author} {\bibfnamefont
  {P.}~\bibnamefont {M\"uller}}, \bibinfo {author} {\bibfnamefont
  {O.}~\bibnamefont {Naviliat-Cuncic}}, \bibinfo {author} {\bibfnamefont
  {J.}~\bibnamefont {Pedersen}}, \bibinfo {author} {\bibfnamefont
  {E.}~\bibnamefont {Smith}}, \bibinfo {author} {\bibfnamefont
  {M.}~\bibnamefont {Sternberg}}, \bibinfo {author} {\bibfnamefont {D.~W.}\
  \bibnamefont {Storm}}, \bibinfo {author} {\bibfnamefont {H.~E.}\ \bibnamefont
  {Swanson}}, \bibinfo {author} {\bibfnamefont {F.}~\bibnamefont {Wauters}},\
  and\ \bibinfo {author} {\bibfnamefont {D.}~\bibnamefont {Zumwalt}},\ }\href
  {https://doi.org/10.1103/PhysRevA.96.053411} {\bibfield  {journal} {\bibinfo
  {journal} {Phys. Rev. A}\ }\textbf {\bibinfo {volume} {96}},\ \bibinfo
  {pages} {053411} (\bibinfo {year} {2017})}\BibitemShut {NoStop}%
\bibitem [{\citenamefont {Bagdasarova}(2019)}]{Bag:2019}%
  \BibitemOpen
  \bibfield  {author} {\bibinfo {author} {\bibfnamefont {Y.}~\bibnamefont
  {Bagdasarova}},\ }\emph {\bibinfo {title} {A measurement of the $e-{\bar
  \nu}_e$ angular correlation coefficient in the decay of $^6{\rm He}$}},\
  \href@noop {} {Ph.D. thesis},\ \bibinfo  {school} {University of Washington}
  (\bibinfo {year} {2019})\BibitemShut {NoStop}%
\bibitem [{\citenamefont {Hong}(2016)}]{Hon:2016b}%
  \BibitemOpen
  \bibfield  {author} {\bibinfo {author} {\bibfnamefont {R.}~\bibnamefont
  {Hong}},\ }\emph {\bibinfo {title} {Developments for a measurement of the
  $\beta-\nu$ correlation and determination of the recoil charge-state
  distribution in $^6{\rm He}$ decay}},\ \href@noop {} {Ph.D. thesis},\
  \bibinfo  {school} {University of Washington} (\bibinfo {year}
  {2016})\BibitemShut {NoStop}%
\bibitem [{\citenamefont {James F.~Ziegler}(2010)}]{Zie:2010}%
  \BibitemOpen
  \bibfield  {author} {\bibinfo {author} {\bibfnamefont {J.~B.}\ \bibnamefont
  {James F.~Ziegler}, \bibfnamefont {M.D.~Ziegler}},\ }\href@noop {} {\bibfield
   {journal} {\bibinfo  {journal} {Nuclear Instruments and Methods in Physics
  Research Section B: Beam Interactions with Materials and Atoms}\ }\textbf
  {\bibinfo {volume} {268}},\ \bibinfo {pages} {1818} (\bibinfo {year}
  {2010})}\BibitemShut {NoStop}%
\bibitem [{SRI()}]{SRIM2013}%
  \BibitemOpen
  \href@noop {} {}\bibinfo {howpublished} {SRIM-2013 code available from
  http://www.srim.org/}\BibitemShut {NoStop}%
\bibitem [{\citenamefont {Calaprice}(1975)}]{PhysRevC.12.2016}%
  \BibitemOpen
  \bibfield  {author} {\bibinfo {author} {\bibfnamefont {F.~P.}\ \bibnamefont
  {Calaprice}},\ }\href {https://doi.org/10.1103/PhysRevC.12.2016} {\bibfield
  {journal} {\bibinfo  {journal} {Phys. Rev. C}\ }\textbf {\bibinfo {volume}
  {12}},\ \bibinfo {pages} {2016} (\bibinfo {year} {1975})}\BibitemShut
  {NoStop}%
\bibitem [{\citenamefont {Glück}(1998)}]{Glu:1998}%
  \BibitemOpen
  \bibfield  {author} {\bibinfo {author} {\bibfnamefont {F.}~\bibnamefont
  {Glück}},\ }\href
  {https://doi.org/https://doi.org/10.1016/S0375-9474(97)00643-X} {\bibfield
  {journal} {\bibinfo  {journal} {Nuclear Physics A}\ }\textbf {\bibinfo
  {volume} {628}},\ \bibinfo {pages} {493} (\bibinfo {year}
  {1998})}\BibitemShut {NoStop}%
\bibitem [{COM()}]{COMSOL}%
  \BibitemOpen
  \href@noop {} {}\bibinfo {howpublished} {COMSOL Multiphysics$^{\rm{TM}}$ v.
  5.6. www.comsol.com. COMSOL AB, Stockholm, Sweden.}\BibitemShut {Stop}%
\bibitem [{Note1()}]{Note1}%
  \BibitemOpen
  \bibinfo {note} {The particular values were chosen for convenience in
  optimizing the Monte Carlo calculations, but do not affect the
  results.}\BibitemShut {Stop}%
\bibitem [{\citenamefont {Fenker}\ \emph {et~al.}(2018)\citenamefont {Fenker},
  \citenamefont {Gorelov}, \citenamefont {Melconian}, \citenamefont {Behr},
  \citenamefont {Anholm}, \citenamefont {Ashery}, \citenamefont {Behling},
  \citenamefont {Cohen}, \citenamefont {Craiciu}, \citenamefont {Gwinner},
  \citenamefont {McNeil}, \citenamefont {Mehlman}, \citenamefont {Olchanski},
  \citenamefont {Shidling}, \citenamefont {Smale},\ and\ \citenamefont
  {Warner}}]{PhysRevLett.120.062502}%
  \BibitemOpen
  \bibfield  {author} {\bibinfo {author} {\bibfnamefont {B.}~\bibnamefont
  {Fenker}}, \bibinfo {author} {\bibfnamefont {A.}~\bibnamefont {Gorelov}},
  \bibinfo {author} {\bibfnamefont {D.}~\bibnamefont {Melconian}}, \bibinfo
  {author} {\bibfnamefont {J.~A.}\ \bibnamefont {Behr}}, \bibinfo {author}
  {\bibfnamefont {M.}~\bibnamefont {Anholm}}, \bibinfo {author} {\bibfnamefont
  {D.}~\bibnamefont {Ashery}}, \bibinfo {author} {\bibfnamefont {R.~S.}\
  \bibnamefont {Behling}}, \bibinfo {author} {\bibfnamefont {I.}~\bibnamefont
  {Cohen}}, \bibinfo {author} {\bibfnamefont {I.}~\bibnamefont {Craiciu}},
  \bibinfo {author} {\bibfnamefont {G.}~\bibnamefont {Gwinner}}, \bibinfo
  {author} {\bibfnamefont {J.}~\bibnamefont {McNeil}}, \bibinfo {author}
  {\bibfnamefont {M.}~\bibnamefont {Mehlman}}, \bibinfo {author} {\bibfnamefont
  {K.}~\bibnamefont {Olchanski}}, \bibinfo {author} {\bibfnamefont {P.~D.}\
  \bibnamefont {Shidling}}, \bibinfo {author} {\bibfnamefont {S.}~\bibnamefont
  {Smale}},\ and\ \bibinfo {author} {\bibfnamefont {C.~L.}\ \bibnamefont
  {Warner}},\ }\href {https://doi.org/10.1103/PhysRevLett.120.062502}
  {\bibfield  {journal} {\bibinfo  {journal} {Phys. Rev. Lett.}\ }\textbf
  {\bibinfo {volume} {120}},\ \bibinfo {pages} {062502} (\bibinfo {year}
  {2018})}\BibitemShut {NoStop}%
\bibitem [{\citenamefont {Ohayon}\ \emph {et~al.}(2020)\citenamefont {Ohayon},
  \citenamefont {Rahangdale}, \citenamefont {Parnes}, \citenamefont {Perelman},
  \citenamefont {Heber},\ and\ \citenamefont {Ron}}]{PhysRevC.101.035501}%
  \BibitemOpen
  \bibfield  {author} {\bibinfo {author} {\bibfnamefont {B.}~\bibnamefont
  {Ohayon}}, \bibinfo {author} {\bibfnamefont {H.}~\bibnamefont {Rahangdale}},
  \bibinfo {author} {\bibfnamefont {E.}~\bibnamefont {Parnes}}, \bibinfo
  {author} {\bibfnamefont {G.}~\bibnamefont {Perelman}}, \bibinfo {author}
  {\bibfnamefont {O.}~\bibnamefont {Heber}},\ and\ \bibinfo {author}
  {\bibfnamefont {G.}~\bibnamefont {Ron}},\ }\href
  {https://doi.org/10.1103/PhysRevC.101.035501} {\bibfield  {journal} {\bibinfo
   {journal} {Phys. Rev. C}\ }\textbf {\bibinfo {volume} {101}},\ \bibinfo
  {pages} {035501} (\bibinfo {year} {2020})}\BibitemShut {NoStop}%
\bibitem [{\citenamefont {Fabian}\ \emph {et~al.}(2018)\citenamefont {Fabian},
  \citenamefont {Fl\'echard}, \citenamefont {Pons}, \citenamefont {Li\'enard},
  \citenamefont {Ban}, \citenamefont {Breitenfeldt}, \citenamefont {Couratin},
  \citenamefont {Delahaye}, \citenamefont {Durand}, \citenamefont {Finlay},
  \citenamefont {Guillon}, \citenamefont {Lemi\`ere}, \citenamefont {Mauger},
  \citenamefont {M\'ery}, \citenamefont {Naviliat-Cuncic}, \citenamefont
  {Porobic}, \citenamefont {Qu\'em\'ener}, \citenamefont {Severijns},\ and\
  \citenamefont {Thomas}}]{PhysRevA.97.023402}%
  \BibitemOpen
  \bibfield  {author} {\bibinfo {author} {\bibfnamefont {X.}~\bibnamefont
  {Fabian}}, \bibinfo {author} {\bibfnamefont {X.}~\bibnamefont {Fl\'echard}},
  \bibinfo {author} {\bibfnamefont {B.}~\bibnamefont {Pons}}, \bibinfo {author}
  {\bibfnamefont {E.}~\bibnamefont {Li\'enard}}, \bibinfo {author}
  {\bibfnamefont {G.}~\bibnamefont {Ban}}, \bibinfo {author} {\bibfnamefont
  {M.}~\bibnamefont {Breitenfeldt}}, \bibinfo {author} {\bibfnamefont
  {C.}~\bibnamefont {Couratin}}, \bibinfo {author} {\bibfnamefont
  {P.}~\bibnamefont {Delahaye}}, \bibinfo {author} {\bibfnamefont
  {D.}~\bibnamefont {Durand}}, \bibinfo {author} {\bibfnamefont
  {P.}~\bibnamefont {Finlay}}, \bibinfo {author} {\bibfnamefont
  {B.}~\bibnamefont {Guillon}}, \bibinfo {author} {\bibfnamefont
  {Y.}~\bibnamefont {Lemi\`ere}}, \bibinfo {author} {\bibfnamefont
  {F.}~\bibnamefont {Mauger}}, \bibinfo {author} {\bibfnamefont
  {A.}~\bibnamefont {M\'ery}}, \bibinfo {author} {\bibfnamefont
  {O.}~\bibnamefont {Naviliat-Cuncic}}, \bibinfo {author} {\bibfnamefont
  {T.}~\bibnamefont {Porobic}}, \bibinfo {author} {\bibfnamefont
  {G.}~\bibnamefont {Qu\'em\'ener}}, \bibinfo {author} {\bibfnamefont
  {N.}~\bibnamefont {Severijns}},\ and\ \bibinfo {author} {\bibfnamefont
  {J.-C.}\ \bibnamefont {Thomas}},\ }\href
  {https://doi.org/10.1103/PhysRevA.97.023402} {\bibfield  {journal} {\bibinfo
  {journal} {Phys. Rev. A}\ }\textbf {\bibinfo {volume} {97}},\ \bibinfo
  {pages} {023402} (\bibinfo {year} {2018})}\BibitemShut {NoStop}%
\bibitem [{\citenamefont {Wilson}\ \emph {et~al.}(2021)\citenamefont {Wilson},
  \citenamefont {Nagel}, \citenamefont {Marley}, \citenamefont {Scielzo},
  \citenamefont {Aprahamian}, \citenamefont {Clark}, \citenamefont
  {Czeszumska}, \citenamefont {Savard}, \citenamefont {Siegl},\ and\
  \citenamefont {Wang}}]{WILSON2021165806}%
  \BibitemOpen
  \bibfield  {author} {\bibinfo {author} {\bibfnamefont {G.}~\bibnamefont
  {Wilson}}, \bibinfo {author} {\bibfnamefont {T.}~\bibnamefont {Nagel}},
  \bibinfo {author} {\bibfnamefont {S.}~\bibnamefont {Marley}}, \bibinfo
  {author} {\bibfnamefont {N.}~\bibnamefont {Scielzo}}, \bibinfo {author}
  {\bibfnamefont {A.}~\bibnamefont {Aprahamian}}, \bibinfo {author}
  {\bibfnamefont {J.}~\bibnamefont {Clark}}, \bibinfo {author} {\bibfnamefont
  {A.}~\bibnamefont {Czeszumska}}, \bibinfo {author} {\bibfnamefont
  {G.}~\bibnamefont {Savard}}, \bibinfo {author} {\bibfnamefont
  {K.}~\bibnamefont {Siegl}},\ and\ \bibinfo {author} {\bibfnamefont
  {B.}~\bibnamefont {Wang}},\ }\href
  {https://doi.org/https://doi.org/10.1016/j.nima.2021.165806} {\bibfield
  {journal} {\bibinfo  {journal} {Nuclear Instruments and Methods in Physics
  Research Section A: Accelerators, Spectrometers, Detectors and Associated
  Equipment}\ }\textbf {\bibinfo {volume} {1017}},\ \bibinfo {pages} {165806}
  (\bibinfo {year} {2021})}\BibitemShut {NoStop}%
\bibitem [{\citenamefont {Martoff}\ \emph {et~al.}(2021)\citenamefont
  {Martoff}, \citenamefont {Granato}, \citenamefont {Palmaccio}, \citenamefont
  {Yu}, \citenamefont {Smith}, \citenamefont {Hudson}, \citenamefont
  {Hamilton}, \citenamefont {Schneider}, \citenamefont {Chang}, \citenamefont
  {Renshaw}, \citenamefont {Malatino}, \citenamefont {Meyers},\ and\
  \citenamefont {Lamichhane}}]{HUNTER:2021}%
  \BibitemOpen
  \bibfield  {author} {\bibinfo {author} {\bibfnamefont {C.}~\bibnamefont
  {Martoff}}, \bibinfo {author} {\bibfnamefont {F.}~\bibnamefont {Granato}},
  \bibinfo {author} {\bibfnamefont {V.}~\bibnamefont {Palmaccio}}, \bibinfo
  {author} {\bibfnamefont {X.}~\bibnamefont {Yu}}, \bibinfo {author}
  {\bibfnamefont {P.}~\bibnamefont {Smith}}, \bibinfo {author} {\bibfnamefont
  {E.}~\bibnamefont {Hudson}}, \bibinfo {author} {\bibfnamefont
  {P.}~\bibnamefont {Hamilton}}, \bibinfo {author} {\bibfnamefont
  {C.}~\bibnamefont {Schneider}}, \bibinfo {author} {\bibfnamefont
  {E.}~\bibnamefont {Chang}}, \bibinfo {author} {\bibfnamefont
  {A.}~\bibnamefont {Renshaw}}, \bibinfo {author} {\bibfnamefont
  {F.}~\bibnamefont {Malatino}}, \bibinfo {author} {\bibfnamefont
  {P.}~\bibnamefont {Meyers}},\ and\ \bibinfo {author} {\bibfnamefont
  {B.}~\bibnamefont {Lamichhane}},\ }\href
  {https://doi.org/10.1088/2058-9565/abdb9b} {\bibfield  {journal} {\bibinfo
  {journal} {Quantum Sci. Technol.}\ }\textbf {\bibinfo {volume} {6}},\
  \bibinfo {pages} {024008} (\bibinfo {year} {2021})}\BibitemShut {NoStop}%
\end{thebibliography}%
\end{document}